\documentclass{article}
\usepackage{spconfa4,amsmath,graphicx}
\usepackage{textcomp}
\usepackage{stfloats} 
\usepackage{url}
\usepackage{verbatim}
\usepackage{graphicx}
\usepackage{cite}
\usepackage{comment}
\usepackage{booktabs}
\usepackage{tabularx}
\usepackage{longtable}
\usepackage{multirow}
\usepackage{subcaption}
\usepackage{colortbl}
\usepackage{amssymb}
\usepackage{xcolor}
\usepackage{pifont}

\newcommand{\xmark}{\ding{55}}%
\newcommand{\cmark}{\ding{51}}%

\title{Ontology-based Target Sound Extraction} 
\name{Carlos Hernandez-Olivan, Marc Delcroix, Tsubasa Ochiai, Naohiro Tawara, Shoko Araki }
\address{NTT Inc., Japan
}
\begin{document}
%
\maketitle
\begingroup
\renewcommand\thefootnote{}\footnotetext{\scriptsize © 2026 IEEE. Personal use of this material is permitted. Permission from IEEE must be obtained for all other uses, in any current or future media, including reprinting/republishing this material for advertising or promotional purposes, creating new collective works, for resale or redistribution to servers or lists, or reuse of any copyrighted component of this work in other works.}
\endgroup

\begin{abstract} 
Target sound extraction (TSE) aims to isolate a sound source of interest from a mixture, given a semantic query. Existing TSE systems are conditioned on fixed class representations tied to individual sound categories, limiting their ability to handle the hierarchical relationships that naturally organize environmental sounds. In this paper, we introduce ontology-based TSE, a new task formulation in which a single model extracts sounds queried at any level of a sound ontology, from fine-grained leaf classes such as \textit{cat} and \textit{dog} to high-level categories such as \textit{animal}. We propose a learnable class embedding table defined over all nodes of an AudioSet-derived ontology, regularized with a Cophenetic Correlation Coefficient (CPCC) loss that aligns embedding distances with shortest-path distances in the ontology tree. Our experiments across different approaches show the benefit of considering the ontology structure when training TSE systems.
\end{abstract}
\section{Introduction}
\label{sec:intro}

The human auditory system is remarkably capable of 
selectively attending to sounds of interest within 
complex acoustic environments, a phenomenon commonly 
referred to as the cocktail party 
effect~\cite{cherry1953some}. Replicating this capability in machines has motivated research in computational auditory scene 
analysis~\cite{bregman1994auditory}, and more recently, target sound extraction (TSE)~\cite{ochiai2020listen,
waveformer}, which aims to isolate a source of interest from a mixture given a semantic or acoustic query~\cite{soundbeamm2d,kong2025universal,LiQCWYLQZ23,liu2024separate,kilgour2022text}. In many practical applications, the system must extract sounds at different levels of semantic granularity. Environmental sounds are naturally organized into hierarchical taxonomies, or \textit{ontologies}, 
where broad categories such as \textit{human sounds} group finer-grained classes such as \textit{voice}, and even more specific classes such as \textit{speech} or \textit{laugh}. A flexible TSE system should be able to handle queries at any level of such hierarchy, from leaf to parent classes, using a single model.

\begin{figure}[!t]
    \centering
    \includegraphics[width=.97\linewidth]{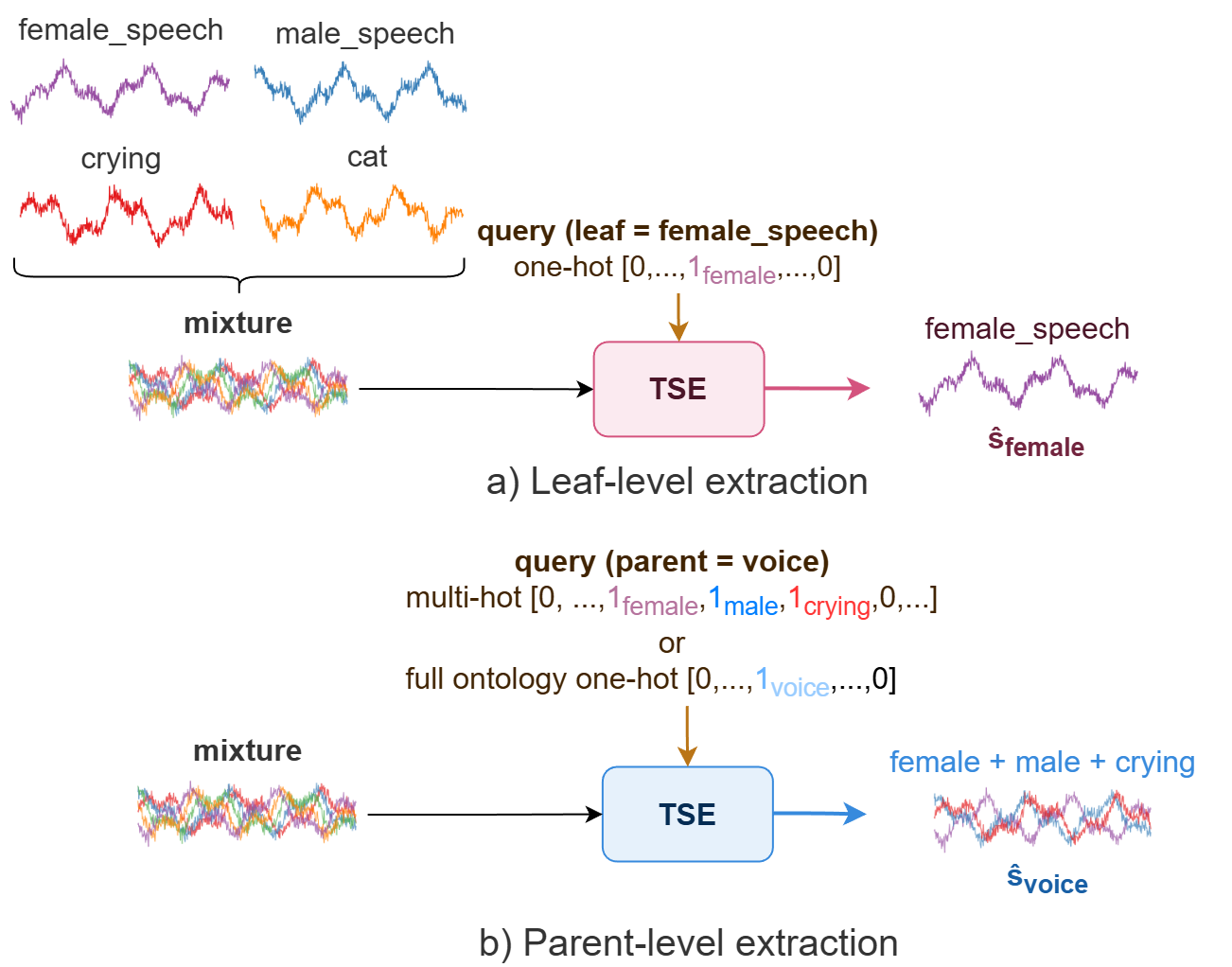}
    \caption{Our proposed Ontology-based TSE. Leaf level or parent classes are extracted with multi-hot or one-hot queries that contain the full ontology nodes. 
    }
    \label{fig:scheme}
\end{figure}

Existing TSE systems, however, typically learn to perform extraction at a fixed sound class granularity. Class-conditioned systems~\cite{LiQCWYLQZ23,waveformer} condition extraction on individual sound class labels, 
with no explicit relationship between semantically 
related categories. Kong et al.~\cite{kong2025universal} 
propose to combine TSE with automatic sound classification, detecting active classes in the mixture and extracting each one individually, then grouping results according to the AudioSet 
ontology~\cite{audioset}. While effective, this approach requires a separate detection step plus one forward pass per active leaf class, and does not consider the ontology structure during TSE training, 
missing the opportunity to learn a better-structured embedding space that could directly benefit extraction. An alternative direction to perform hierarchical extraction, in this case proposed for sound separation, is to explicitly shape the embedding space to reflect semantic hierarchies~\cite{petermann2023hyperbolic} in the hyperbolical space. However, this method does not provide an explicit and measurable alignment between embedding geometry and ontology tree structure.

In this work, we propose a TSE model that can extract sounds at any ontology level with a single forward pass, without requiring a separate detection stage.
We encode the hierarchical structure of 
a sound ontology directly into the conditioning 
embedding space of a TSE model 
(Figure~\ref{fig:scheme}). We investigate two ontology-aware conditioning strategies: multi-hot (i.e., to extract \textit{voice}, the multi-hot vector has the entries for \textit{speech} and \textit{laugh} set to ones, which activates multiple leaf labels at once), and full-ontology one-hot, which maps queries to a single unique leaf node (i.e., \textit{speech}, \textit{laugh}, \textit{voice} and \textit{human sounds} are all considered as different classes with different one-hot vectors). In addition, we introduce a Cophenetic Correlation Coefficient (CPCC) loss~\cite{zeng2022learning} to regularize the embedding space, encouraging semantically related classes to have similar representations. The CPCC loss, which was first proposed for image classification, ensures that pairwise distances in a learned latent space reflect shortest-path distances in a reference tree. 
Our contributions are: (i) a conditioning framework for ontology-based TSE that enables single-pass extraction at any ontology level through CPCC-regularized class embeddings; (ii) a comparison between different conditioning queries with and without CPCC regularization.

\begin{figure*}[h]
    \centering
    \includegraphics[width=\textwidth]{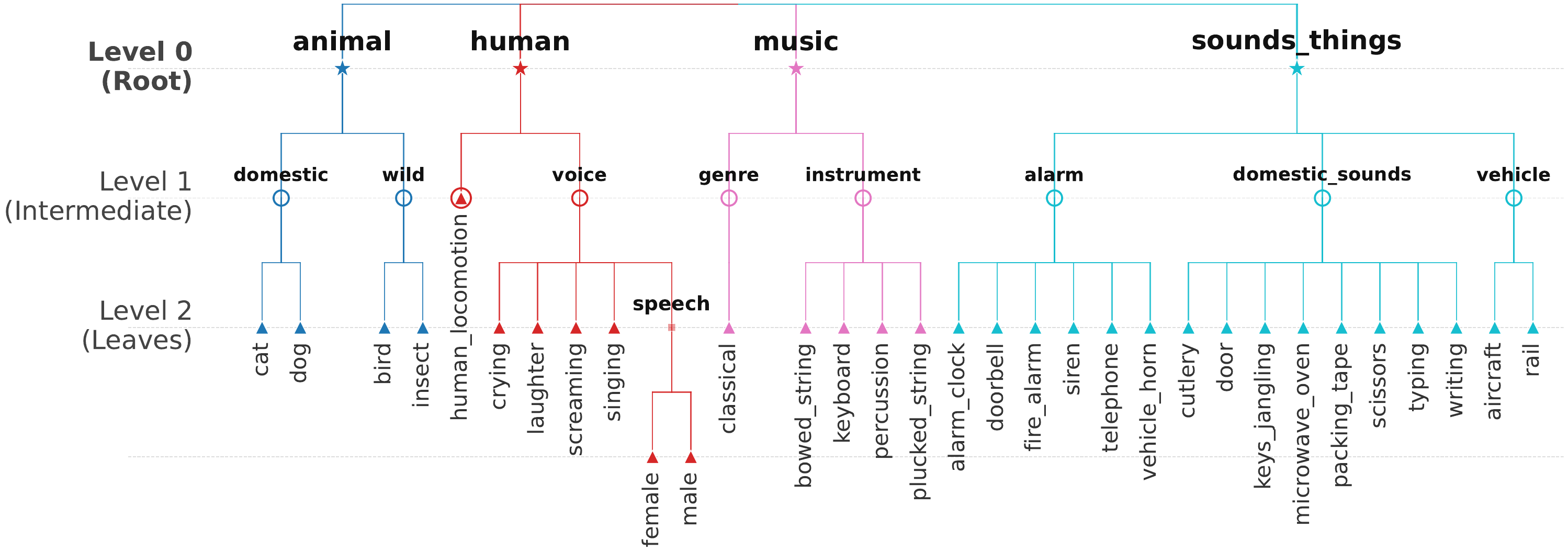}
    \caption{Our ontology derived from AudioSet. Stars, circles, and triangles represent roots, intermediates, and leaves, respectively. Classes represented with both circles and triangles are used for both intermediate and leave levels evaluation.}
    \label{fig:ontology}
\end{figure*}

\section{Ontology-based TSE}
\label{sec:tse}
Ontology-based TSE aims to extract all sounds belonging to a queried node $v \in \mathcal{V}$ from a mixture $\mathbf{x}$, at any ontology level, using a single model and a single forward pass. We show our AudioSet-derived ontology in Fig.~\ref{fig:ontology}. The ontology is structured in three main levels: Root or Level 0, which corresponds to broad sound categories, Intermediate or Level 1, which contains more specific sound classes, and leaves, which correspond to the finest-grained level. We train our models within this ontology and evaluate across three ontology levels as shown in Figure~\ref{fig:ontology} with stars, circles, and triangles. 

\subsection{TSE Architecture}
In this research, we focus on class-based TSE, where the target sound class is represented by an embedding vector derived from a one-hot or multi-hot vector through an embedding layer~\cite{ochiai2020listen}. Let $\mathcal{C}$ be a set of sound classes with $|\mathcal{C}|$ classes. For a target class $c \in \mathcal{C}$,  the embedding vector is: $\mathbf{e}_c = \mathbf{W}^\top \mathbf{o}_c$, where $\mathbf{o}_c \in \{0,1\}^{|\mathcal{C}|}$ is a one-hot vector and $\mathbf{W} \in \mathbb{R}^{|\mathcal{C}| \times D}$ is a learnable embedding matrix with embedding dimension $D$. The system's output is then computed as:
\begin{equation}
    \mathbf{\hat{s}}_c = \text{TSE}(\mathbf{x}, \mathbf{W}^\top \mathbf{o}_c) .
\end{equation}

Our system builds upon the SoundBeam architecture~\cite{soundbeam}, which follows an encoder-masker-decoder paradigm operating in the time domain. A 1D-convolutional encoder maps the input mixture $\mathbf{x} \in \mathbb{R}^T$ into a latent representation where $T$ is the signal length in samples. A TasNet-based masker~\cite{tasnet} computes a multiplicative mask conditioned on a query embedding $\mathbf{e} \in \mathbb{R}^D$, and a decoder reconstructs the estimated source $\hat{\mathbf{s}} \in \mathbb{R}^T$. The conditioning is done by multiplying the latent representation of the mixture in the masker by the conditioning embedding vector. 
We use the same architecture as described in~\cite{soundbeam}.

\subsection{Ontology-Aware Conditioning}

Let $\mathcal{T}$ be a rooted tree representing the sound class ontology, where each node $v \in \mathcal{V}$ corresponds to a sound category at a specific level of granularity (see Fig.~\ref{fig:ontology}). Leaf nodes represent fine-grained classes (e.g., \textit{female speech}, \textit{dog}), while internal nodes group related classes into broader categories (e.g., \textit{voice}, \textit{domestic animal}). We denote by $\mathcal{L}(v) \subset \mathcal{V}$ the set of leaf nodes in the subtree rooted at $v$, and by $\mathcal{L}$ the set of all leaf nodes.
We investigate three conditioning strategies, i.e., a baseline one-hot TSE system with no ontology structure (Section~\ref{sec:case1}), a multi-hot (Section~\ref{sec:case2}), and one-hot of the full ontology inputs (Section~\ref{sec:case3}).

\subsubsection{One-Hot Leaf Extraction and Aggregation}
\label{sec:case1}

For the baseline, we consider performing TSE for all the leaves $\mathcal{L}_v$ of each root sound class $v$, and summing the extracted signals:
\begin{equation}
    \hat{s}_v(\mathbf{x}) = \sum_{\ell \in 
    \mathcal{L}(v)}
    \mathrm{TSE}\!\left(\mathbf{x};\, 
    \mathbf{W}^\top \mathbf{o}_\ell\right),
    \label{eq:case1_parent}
\end{equation}
where $\mathbf{o}_\ell$ is the one-hot vector corresponding to the leaf $\ell$ of node $v$. We train the model to extract only the active leaves in the mixture. At inference, we extract all leaf descendants regardless of their presence in the mixture, reflecting a realistic scenario where active classes are unknown. This aggregation strategy is conceptually similar to~\cite{kong2025universal}, which independently extracts each automatically detected leaf class of the desired node according to the AudioSet ontology. 
The key difference is that Kong et al.~\cite{kong2025universal} use a pretrained audio classifier to identify active leaves, whereas we do not use it at the cost of extracting absent classes.

\subsubsection{Multi-Hot Hierarchical Conditioning} \label{sec:case2}

A one-hot input vector of only leaf nodes does not contain ontology information, and it does not enable single-pass extraction at arbitrary ontology levels. Therefore, we extend the
conditioning to a multi-hot vector~\cite{ochiai2020listen} that activates leaf descendants of the queried node.
For any node $v \in \mathcal{V}$:
\begin{equation}
    \mathbf{e}^{\text{m-hot}}_v = \mathbf{W}^\top 
    \mathbf{m}_v,
    \label{eq:case2}
\end{equation}
where $\mathbf{m}_v \in \{0,1\}^{|\mathcal{L}|}$ is a multi-hot vector whose $\ell$-th entry is 1 if leaf $\ell \in \mathcal{L}$ is a descendant of $v$ active or not in the mixture (i.e., $\ell \in \mathcal{L}(v)$), and 0 otherwise, and $\mathbf{W} \in \mathbb{R}^{|\mathcal{L}| \times D}$. This exact same multi-hot query is used during both training and inference. The reconstruction target is the sum of isolated signals of all the descendants of $v$ present in the mixture. 
This approach is computationally more efficient than the approach in Section \ref{sec:case1}, as it only requires a single pass and implicitly learns the ontology structure by computing the parent embeddings as the sum of its children's embeddings.

\subsubsection{Full-Ontology One-Hot Conditioning}
\label{sec:case3}

In previous formulations, we defined the embedding matrix $\mathbf{W}$ only over leaf nodes. Consequently, internal nodes do not have dedicated embeddings: they are handled either by aggregating descendant leaf extractions (Section~\ref{sec:case1}) or by representing them as multi-hot combinations over descendant leaves (Section~\ref{sec:case2}). We also explore an alternative, where we extend the input space to cover \emph{all} $|\mathcal{V}|$ nodes, leaves and internal ancestors, assigning each node its own independent embedding vector. For a queried node $v \in \mathcal{V}$:
\begin{equation}
    \mathbf{e}_v = \mathbf{\overline{W}}^\top \mathbf{o}_v,
    \label{eq:case3}
\end{equation}
where $\mathbf{o}_v \in \{0,1\}^{|\mathcal{V}|}$ is the one-hot for node $v$, $\mathbf{\overline{W}} \in \mathbb{R}^{|\mathcal{V}| \times D}$. Using an embedding per ontology node allows ancestor categories to be modeled as independent semantic concepts rather than fixed combinations of leaf embeddings.

\subsection{Training Strategy and Loss Function} \label{sec:cpcc}

For training the multi-hot and full-ontology one-hot systems, we use the following strategy: at each batch step, one node per ontology level is selected, and the reconstruction target is the sum of isolated signals of its active leaf descendants. We use the negative Scale Invariant Signal-to-Noise Ratio (SI-SNR) as the reconstruction loss $\mathcal{L}_{\text{rec}}$. In addition, we introduce a CPCC loss to regularize the embedding space in the TSE task.


One way to impose the ontology structure in the embedding space is to make embeddings of leaves with the same parent close to each other and different from other sound categories. We can realize this by enforcing that the embedding space reflects the structure of $\mathcal{T}$ by minimizing the CPCC loss~\cite{zeng2022learning} defined as the negative Pearson correlation between tree and embedding distances over all node pairs: 

\begin{equation}
\resizebox{\columnwidth}{!}{
    $ \displaystyle
    \mathcal{L}_{\text{CPCC}} = -\frac{
        \sum_{u < v}
            \bigl(d_{\mathcal{T}}(u,v) - \bar{d}_{\mathcal{T}}\bigr)
            \bigl(d_{\mathcal{Z}}(u,v) - \bar{d}_{\mathcal{Z}}\bigr)
    }{\sqrt{
        \sum_{u < v}\!\bigl(d_{\mathcal{T}}(u,v)-\bar{d}_{\mathcal{T}}\bigr)^2\;
        \sum_{u < v}\!\bigl(d_{\mathcal{Z}}(u,v)-\bar{d}_{\mathcal{Z}}\bigr)^2
    }} ,$
}
\label{eq:cpcc}
\end{equation}
where $d_{\mathcal{T}}(u,v)$ is the tree distance between nodes $u$ and $v$ on $\mathcal{T}$, and $d_{\mathcal{Z}}(u,v) = \|\mathbf{e}_u - \mathbf{e}_v\|_2$ is the Euclidean distance between the embeddings of nodes $u$ and $v$. $\bar{d}_{\mathcal{T}}$ and $\bar{d}_{\mathcal{Z}}$ denote the means of $d_{\mathcal{T}}$ and $d_{\mathcal{Z}}$ over all node pairs, respectively.
Here, the tree distance is defined as the shortest-path length between nodes $u$ and $v$ on $\mathcal{T}$, $d_{\mathcal{T}}(u,v)$; for example, two leaves sharing a direct parent have $d_{\mathcal{T}}=2$, a leaf and its immediate ancestor have $d_{\mathcal{T}}=1$, whereas leaves in different Level-0 branches can reach $d_{\mathcal{T}}$ up to $2\cdot\text{depth}(\mathcal{T})$. 


The total loss when adding the regularization is:
\begin{equation}
      \mathcal{L} = \mathcal{L}_{\text{rec}} + \lambda 
  \mathcal{L}_{\text{CPCC}}
  \end{equation}
where $\lambda$ controls the weight of the hierarchical constraint.
\section{Experiments and Results}
\label{sec:results}

\begin{table}[h]
    \scriptsize
    \centering
    \setlength{\tabcolsep}{2.5pt}
    \caption{Mean SNR / SI-SNR improvements (dB) evaluated on the test set. The unprocessed row reports the SNR/SI-SNR between the mixtures and the targets.}
    \label{tab:tse_results_compact}
    \resizebox{\columnwidth}{!}{
    \begin{tabular}{l@{}ccccccc}
    \toprule
    & \multirow{2}{*}{\textbf{CPCC}} & \multicolumn{2}{c}{\textbf{Root (L0)}} & \multicolumn{2}{c}{\textbf{Interm. (L1)}} & \multicolumn{2}{c}{\textbf{Leaves}}  \\
    \cmidrule(lr){3-4}\cmidrule(lr){5-6}\cmidrule(lr){7-8}
    \textbf{Conditioning} & \textbf{loss} & \textbf{SNRi} & \textbf{SI-SNRi} & \textbf{SNRi} & \textbf{SI-SNRi} & \textbf{SNRi} & \textbf{SI-SNRi} \\
    \midrule
    0) \textit{Unprocessed (SI-SNR abs.)}
        & -
        & -0.69 & -0.69
        & -2.00 & -2.00
        & -5.43 & -5.43 \\
    
    \midrule
    1) Leaf extr. + Aggr. (Oracle) 
    & \xmark & 
        5.57 & 3.79 & 6.50 & 4.22 & 9.04 & 5.42 \\
    2) Leaf extr. + Aggr (All)
        & \xmark & 
        -4.67 & 1.52 & 1.26 & 2.97 & 9.04 & 5.42 \\
\midrule
    3) multi-hot
        & \xmark & 6.93 & 5.57 & 7.62 & 5.93 & 9.24 & 6.12 \\ 
    4) 
        & \cmark & \textbf{7.72} & \textbf{6.43} & \textbf{8.36} & \textbf{6.66} & \textbf{10.21} & \textbf{7.10} \\

   \midrule     
    5) full ontology one-hot
        & \xmark & 5.87 & 4.42 & 6.77 & 4.90 & 8.99 & 5.68 \\ 
    6) 
        & \cmark & 5.52 & 4.75 & 6.44 & 5.12 & 8.78 & 5.73 \\ 

        
    
        
    
        
    \bottomrule
    \end{tabular}
    }
\end{table}

\subsection{Ontology-Annotated Dataset}
Our dataset is synthesized using four public collections: FSD50K~\cite{FSD50K}, LibriSpeech~\cite{librispeech} (for high-quality speech that we substitute from FSD50K), SynthSOD~\cite{synthsod}, and MoisesDB~\cite{moisesdb} (for instrument stems and \textit{classical} genre). To ensure both intra-class variety and inter-class diversity, we use two-stage hierarchical sampling based on the ontology. First, we perform \textit{intra-subtree sampling} by randomly selecting a parent node at either the root (Level 0) or intermediate level (Level 1) and drawing 2--3 leaf nodes from its descendants. Second, we apply \textit{cross-tree sampling} by selecting a node at the same hierarchical depth but under a different root ancestor (Level 0), from which an additional 2--3 leaf nodes are drawn. This guarantees that each mixture, comprising 3--5 targets with durations between 2.5 and 3.5 seconds, contains at least two root branches. The duration of each mixture is 5 seconds. Individual leaf target sources are mixed at an SNR of 5--15 dB relative to the background. The dataset consists of 48K training, 4K validation, and 8K test mixtures.

\subsection{Experimental setting}
We used the same hyperparameters as in~\cite{soundbeamm2d} with $\lambda=0.1$ for the CPCC loss. We used one 1D-Conv and 1D-DeConv layer for the encoder and decoder, respectively, with a kernel size of 16 and a stride of 8. For the extractor, we used 512 filters, 8 blocks, and a global layer norm. For the embedding module, we used 2 fully connected layers with ReLU non-linear layers and layer normalization following \cite{semhear}. We set the embedding size $D=256$. We train all our systems with a batch size of 8 and a learning rate of 1e--5. 

\subsection{Discussion}
Table \ref{tab:tse_results_compact} shows the extraction performance for different ontology levels using the baseline leaves extraction and aggregation method (``leaf extr. + Aggr.’’) with oracle sound event detection (system 1, ``Oracle'') and when performing extraction for all leaves belonging to a parent (system 2, ``All''). We compare the baseline with our proposed multi-hot hierarchical conditioning (systems 3-4, ``multi-hot’’) and full ontology one-hot conditioning (systems 5-6, ``full ontology one-hot’’). The first line shows the absolute SNR/SI-SNR between the mixture and the targets. Given the task setting, root-level extraction has a higher input SNR because there is more than one event from the same root in a mixture, and since the target at L0 comprises the sum of all active descendants under that root, the target at root level represents a larger fraction of the mixture energy than a single leaf-level target. 

From the table, we observe that the baseline (system 2, ``All'') can perform leaf-level extraction but performs very poorly at intermediate or root-level nodes of the ontology. This is because our baseline aggregates the extracted signals from all leaf nodes of a parent, even for leaves corresponding to acoustic events that are inactive in the mixture. For such inactive events, the baseline extraction may perform poorly. This behavior can be mitigated by using a sound event detection system to identify active sound classes, as in \cite{kong2025universal}, at the cost of requiring a separate model. Assuming an oracle sound event detection (system 1), the oracle baseline can be seen as an upper bound of the performance of the approach proposed in~\cite{kong2025universal}.

Comparing our proposed schemes for ontology-level extraction (systems 3 to 6 in Table~\ref{tab:tse_results_compact}), multi-hot extraction outperforms both the baseline and the full-ontology one-hot cases by more than 1 dB for root- and intermediate-level extraction. 

Overall, our proposed ontology-based multi-hot extraction with CPCC regularization achieves stable extraction performance at different ontology levels. These results demonstrate that ontology-based target-sound extraction is possible in a single pass, without a separate sound-detection model. Moreover, comparing the leaf-level extraction performance, we observe a 2.4 dB improvement over the baseline, demonstrating that considering the sound ontology hierarchy during training can also be beneficial for leaf-level extraction. A detailed analysis of CPCC sensitivity is left for future work.

\section{Conclusions}
\label{sec:conclusions}
We have introduced an ontology-aware conditioning framework for TSE that enables single-pass extraction at any level of a sound ontology, from fine-grained leaf classes to root categories, through class embeddings whose geometry reflects the AudioSet ontology. Future work includes exploring audio enrollment queries for ontology-based extraction and exploiting sound ontologies for sound event detection and zero-shot separation.

\vspace*{2mm}
\noindent{\bf Acknowledgments:} This work was supported by JST Strategic International Collaborative Research Program (SICORP), Grant Number JPMJSC2306, Japan.

\bibliographystyle{IEEEbib}
\bibliography{refs}

\end{document}